\documentclass[conference]{IEEEtran}
\IEEEoverridecommandlockouts
\usepackage{cite}
\usepackage{amsmath,amssymb,amsfonts}
\usepackage{algorithmic}
\usepackage{graphicx}
\usepackage{textcomp}
\usepackage{xcolor}
\usepackage{comment}
\usepackage{hyperref}
\hypersetup{colorlinks=true,linkcolor=black,urlcolor=black,citecolor=black}
\usepackage{enumitem} %
\usepackage{flushend} %

\def\BibTeX{{\rm B\kern-.05em{\sc i\kern-.025em b}\kern-.08em
    T\kern-.1667em\lower.7ex\hbox{E}\kern-.125emX}}

\newcommand{\ie}{{i.e. }}
\newcommand{\eg}{{e.g. }}

\begin{document}

\begin{titlepage}
\quad\\[1cm]
\makeatother
	{\Huge IEEE Copyright Notice}\\[0.5cm]
	{\large \copyright \ 2023 IEEE. Personal use of this material is permitted. Permission from IEEE must be obtained for all other uses, in any current or future media, including reprinting/republishing this material for advertising or promotional purposes, creating new collective works, for resale or redistribution to servers or lists, or reuse of any copyrighted component of this work in other works. \\[0.5cm]}    
    {\large Cite as:\\[0.1cm]}
    {\large A. Pino, D. Margaria and A. Vesco, “Combining Decentralized IDentifiers with Proof of Membership to Enable Trust in IoT Networks,” 
    in \textit{Proceedings of 33\textsuperscript{rd} International Telecommunication Networks and Applications Conference}, Melbourne, Australia, 2023, pp. 310-317, doi: 10.1109/ITNAC59571.2023.10368540.}
\end{titlepage}

\title{%
Combining Decentralized IDentifiers with Proof of Membership to Enable Trust in IoT Networks
\thanks{%
This work has been developed within the MASTERMINE project (European Mining in the Green and Digital Era, \url{https://www.mastermine-project.eu/}), funded by the European Union under the Horizon Europe framework programme [GA 101091895].}}

\author{\IEEEauthorblockN{Alessandro Pino, Davide Margaria, Andrea Vesco}
\IEEEauthorblockA{%
\textit{LINKS Foundation - Cybersecurity Research Group}\\
Torino 10138, Italy \\
}

}

\maketitle

\begin{abstract}
The Self-Sovereign Identity (SSI) is a decentralized paradigm enabling full control over the data used to build and prove the identity. In Internet of Things networks with security requirements, the Self-Sovereign Identity can play a key role and bring benefits with respect to centralized identity solutions. The challenge is to make the SSI compatible with resource-constraint IoT networks. In line with this objective, the paper proposes and discusses an alternative (mutual) authentication process for IoT nodes under the same administration domain. The main idea is to combine the Decentralized IDentifier (DID)-based verification of private key ownership with the verification of a proof that the DID belongs to an evolving trusted set. The solution is built around the proof of membership notion. The paper analyzes two membership solutions, a novel solution designed by the Authors based on Merkle trees and a second one based on the adaptation of Boneh, Boyen and Shacham (BBS) group signature scheme. The paper concludes with a performance estimation and a comparative analysis.

\end{abstract}

\begin{IEEEkeywords}
Self-Sovereign Identity, Decentralized IDentifiers, Proof of Membership, 
Group Signatures, Merkle Trees, Trust, Internet of Things.
\end{IEEEkeywords}

\section{Introduction}
\label{intro}
\begin{figure}[t]
  \centerline{\includegraphics[width=\columnwidth]{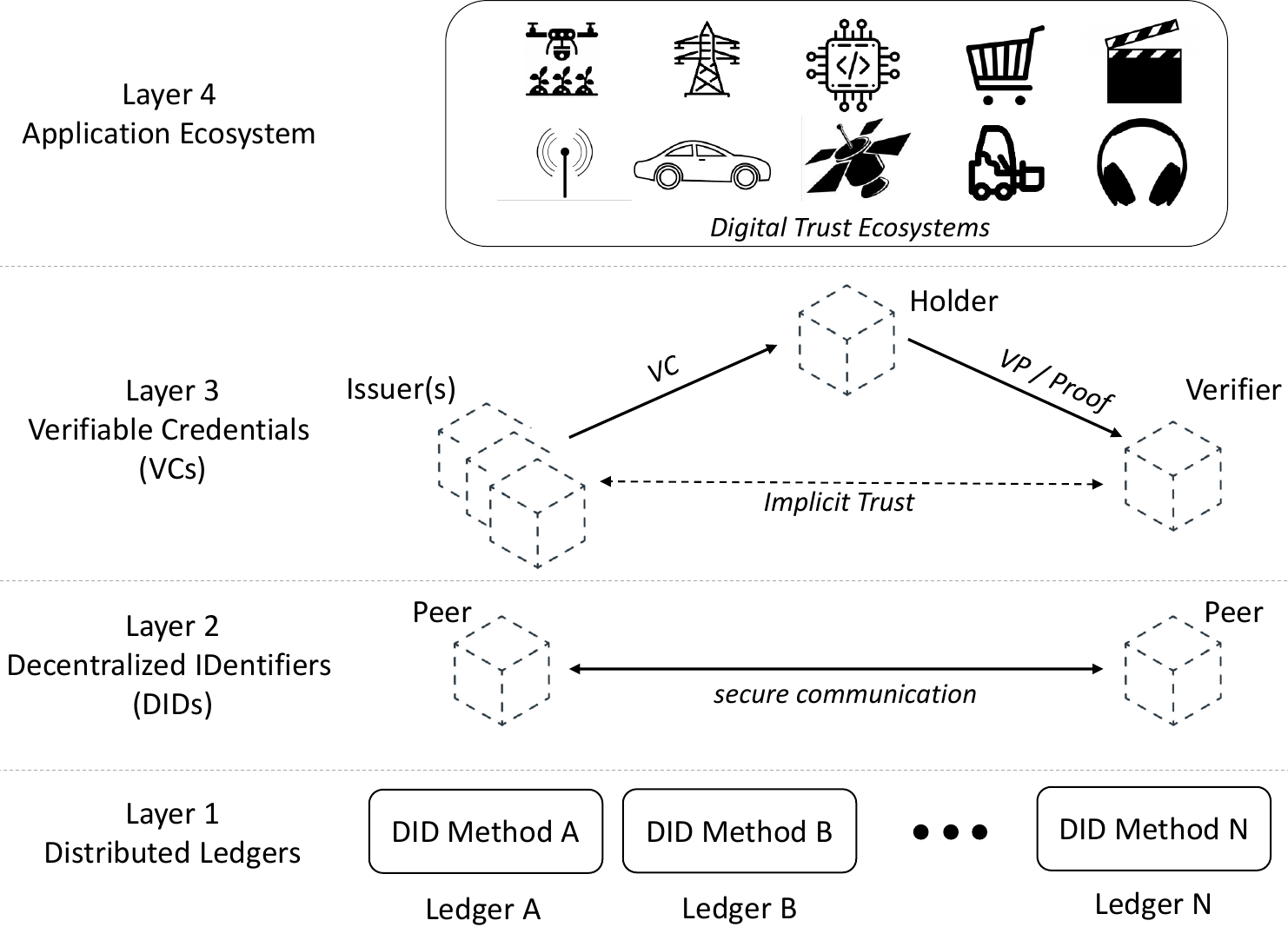}}
  \caption{The Self-Sovereign Identity stack.} 
  \label{ssi}
 \end{figure}

The Self-Sovereign Identity (SSI)~\cite{SSIbook} is a decentralized digital identity paradigm that gives a peer full control over the data it uses to build and to prove its identity. The overall SSI stack, depicted in Fig.~\ref{ssi}, enables a new model for trusted digital interactions.

 The Layer 1 is implemented by means of any Distributed Ledger Technology (DLT) acting as the Root-of-Trust (RoT) for identity data. In fact, DLTs are distributed and immutable means of storage by design~\cite{DLTs}. A Decentralized IDentifier (DID)~\cite{DID} is the new type of globally unique identifier designed to verify a peer. 
 The DID is a Uniform Resource Identifier (URI) of the following form:
 \begin{center}
 \emph{did:method-name:method-specific-id}    
 \end{center}
 where \emph{method-name} is the name of the DID Method used to interact with the DLT and \emph{method-specific-id} is the pointer to the DID Document stored on the DLT, denoted as \emph{index} in this paper for simplicity.

 Thus, DIDs associate a peer with a DID Document~\cite{DID} to enable trustable interactions with it. The DID Method~\cite{DID,DID-registry} is the software implementation used by a peer to interact with the DLT. In accordance with W3C recommendation~\cite{DID}, a DID Method must provide the primitives to:
 
\begin{itemize}

\item \textit{create} a DID, that is, generate an identity key pair ($sk_{id}, pk_{id}$) for authentication purposes, the corresponding DID Document containing the public key of the pair $pk_{id}$ and store the DID Document into the distributed ledger at the \emph{index} pointed by the DID, 
 \item \textit{resolve} a DID, that is, retrieve the DID Document from the \emph{index} on the ledger pointed to by the DID, 
 \item \textit{update} a DID, that is, generate a new key pair ($sk_{id}', pk_{id}'$) and store a new DID Document at the same \emph{index} or at a new \emph{index} if the subject requires changing the DID, 
 and 
 \item \textit{revoke} a DID, that is, provide an immutable evidence on the ledger that a DID has been revoked by the owner. 
 \end{itemize}
 
 The DID Method implementation is ledger-specific and it makes the upper layers independent of the DLT of choice. 

The Layer 2 makes use of DIDs and DID Documents to establish a secure channel between two peers. In principle, both peers prove the ownership of their private key $sk_{id}$ bound to the public key $pk_{id}$ in their DID Document that is stored on the distributed ledger.
While the Layer 2 leverages DID technology (\ie the security foundation of the SSI stack) to begin the authentication procedure, the Layer 3 finalizes it and deals with authorization to services and resources with Verifiable Credentials (VCs)~\cite{VC}. 

A VC is an unforgeable, secure, and machine verifiable digital credential that contains further characteristics of the digital identity of a peer than its key pair ($sk_{id}, pk_{id}$), the DID and the related DID Document.

\emph{The combination of the key pair ($sk_{id}, pk_{id}$), the DID, the corresponding DID Document and at least one VC forms the digital identity in the SSI framework.} 
This composition of the digital identity reflects the decentralized nature of SSI. There is no authority that provides all the components of the identity to a peer, and no authority is able to revoke completely the identity of a peer. Moreover, a peer can enrich its identity with multiple VCs issued by different Issuers.
 
The Layer 3 works in accordance with the Triangle-of-Trust depicted in Fig.~\ref{ssi}.
Three different roles coexist:
\begin{itemize}
  \item \textbf{Holder} is the peer that possesses one or more VCs and that generates a Verifiable Presentation (VP) to request a service or a resource from a Verifier; 
  \item \textbf{Issuer} is the peer that asserts claims about a subject, creates a VC from these claims, and issues the VC to the Holders.
  \item \textbf{Verifier} is the peer that receives a VP from the Holder and verifies the two signatures made by the Issuer on the VC and by the Holder on the VP before granting the access to a service or a resource based on the claims.
\end{itemize}

The VC contains the metadata to describe properties of the credential (\eg context, ID, type, Issuer of the VC, issuance and expiration dates) and most importantly, the DID and the claims about the identity of the peer in the \verb+credentialSubject+ field.  

The Issuer signs the VC to make it an unforgeable and verifiable digital document. 
The Holder requests access to services and resources from the Verifier by presenting a VP. A VP is built as an envelope of the VC. The VC is issued by an Issuer and a signature is made by the Holder with his $sk_{id}$. Issuers are also responsible for VCs revocation for cryptographic integrity and for status change purposes~\cite{VC}.

\begin{figure}[t]
  \centerline{\includegraphics[width=\columnwidth]{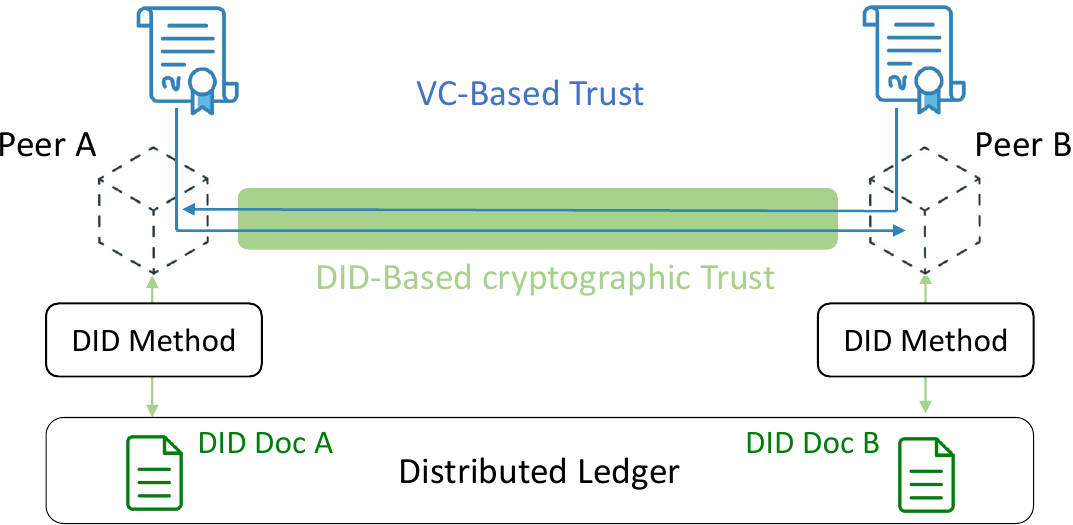}}
  \caption{Mutual authentication between two peers in the SSI framework.} 
  \label{didcom}
 \end{figure}

On top of these three layers, it is possible to build any ecosystem of trustable interactions among peers. The authentication process at the core of Trust between two SSI-aware peers is depicted in Fig.~\ref{didcom}.

In principle, the peers use an (ephemeral) Diffie–Hellman key exchange to build up a confidential channel.
Then, the peers exchange their respective DIDs and prove the possession of the $sk_{id}$ associated to the $pk_{id}$ stored in their corresponding DID Documents.
This verification, in case of success, ends up with a cryptographic trust between the two peers while preventing the passive Man-in-the-Middle (MitM) attack. However, in a permissionless distributed ledger, anyone is entitled to create its own DID, therefore the procedure is still vulnerable to active MitM attack. In fact, the (mutual) authentication takes place only after the peers exchange and verify the respective VCs. At that point, the peers establish a secure communication channel. %

There are Internet of Things (IoT) use cases in which networks of nodes support or make themselves digital infrastructures with security requirements, such as (mutual) authentication, confidentiality, and integrity. The SSI framework can play a key role and bring benefits with respect to centralized identity solutions~\cite{DPKI}. The challenge is to make these solutions compatible with resource constraints~\cite{IoT}. 

With the aim of pursuing this objective, the paper proposes and discusses an alternative (mutual) authentication process for IoT nodes under the same administration domain. 
Finalizing the authentication by means of VC verification at Layer 3 is the most demanding operation of the authentication procedure depicted in Fig.~\ref{didcom} due to the complex data model of VCs~\cite{VC}. 
The proposed alternative is to complement the DID-based verification of the $sk_{id}$ ownership with the verification of a proof that a DID belongs to an evolving trusted set of DIDs (\ie\ the DID has not been created by an adversarial node). 
In other words, the idea is to complement the DID-based verification of the $sk_{id}$ ownership with the verification of a \emph{proof of membership} and avoid the use of VCs.
For the sake of clarity, the membership concept refers to DIDs and not to nodes.

From an implementation point of view, a node combines the DID with the proof of membership and forwards them to the counterpart node that proceeds with the verification of the proof of membership and of the DID ownership (\ie $sk_{id}$ ownership) to complete the authentication procedure.

This paper proposes and analyzes two membership solutions for the purpose of implementing the new authentication procedure: a novel solution based on Merkle trees \cite{MerkleRalphC} designed by the Authors of this paper and 
a second solution built as an adaptation of a well-known and largely used group signature scheme proposed by Boneh, Boyen and Shacham and conventionally referred to as BBS~\cite{BBS}. 

The paper presents and critically reviews the two proposed solutions in four typical operational phases of an IoT network, namely: 

\begin{enumerate}%
\item %
\textbf{Provisioning}: corresponding to the initial setup of the IoT nodes in the network; 
\item %
\textbf{Operation}: corresponding to the operation of the IoT nodes when deployed on the field; 
\item %
\textbf{Secret rotation}: corresponding to the update of node identity keys ($sk_{id},pk_{id}$) and other relevant secret keys; 
\item %
\textbf{Network update}: corresponding to the action of  either adding or removing an IoT node to/from the network.
\end{enumerate}

\section{Membership through Merkle trees} 
\label{sect:MT}
A Merkle tree, also known as a Hash tree, is a data structure that is used to efficiently verify the integrity of large amounts of data. It is named after its inventor, Ralph C. Merkle, who first introduced the concept in a patent filed in 1979 \cite{MerkleRalphC}. 

The Merkle tree is here used to solve the membership problem presented in Section~\ref{intro}. In principle, the DIDs selected by a node must be part of a Merkle tree whose root is considered trusted.

\subsection{Provisioning} 
\label{MerkleProvis}

\begin{figure}[t!]
\centerline{\includegraphics[width=\columnwidth]{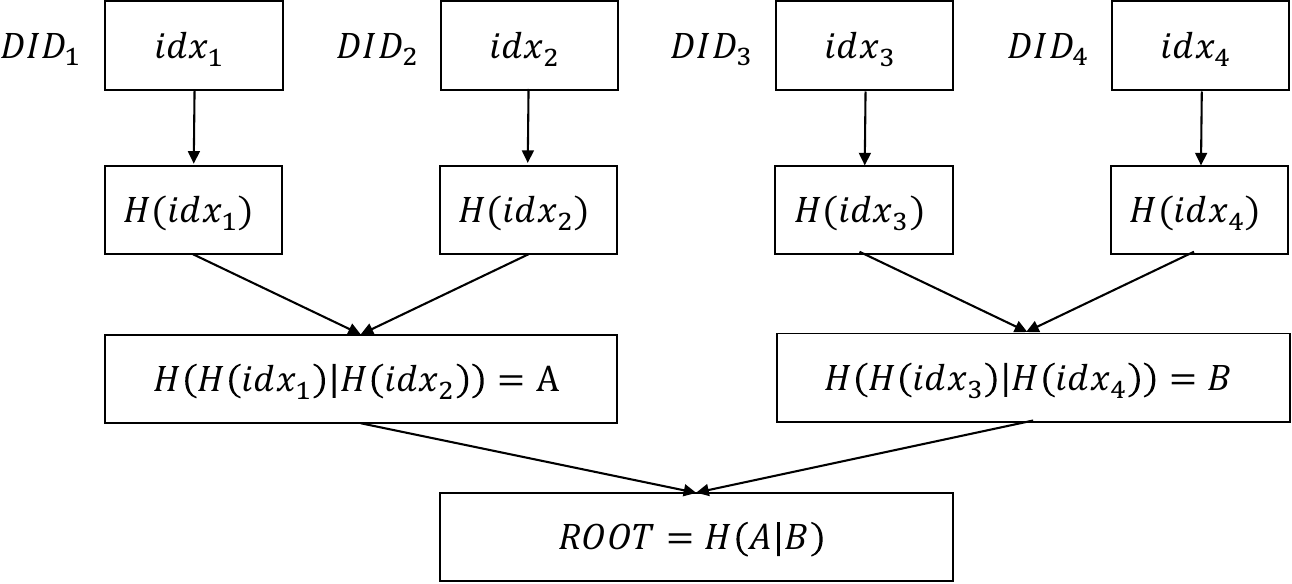}}
\caption{Example of Merkle tree with four Decentralized IDentifiers.}
\label{merkle}
\end{figure}

The provisioning phase consists of the typical configuration procedure, in a secure environment, of each node before their deployment on the field. 

Upon configuration, each node selects a first set of DIDs that will use during the operation phase (\ie defines the indexes $idx_{i}$ of these DIDs). Then, the node generates autonomously its own Merkle tree, as depicted in Fig.~\ref{merkle}.

Basically, the node uses the selected indexes $idx_{i}$ as inputs for a Hash function $H(\cdot)$: the outputs are the leaves of the Merkle tree (\eg  $leaf_{1}=H(idx_{1})$). 
Reminding that every element that is not a leaf is the digest of its child elements (\eg  $parent$ = $H(child_{1}|child_{2})$), the construction consists in hashing the previous two values until only a value remains; this value is called $ROOT$. 

In a possible design, all the leaves of the Merkle tree can be calculated starting from a single master secret $S$. A HMAC-based Extract-and-Expand Key Derivation Function~\cite{HKDFdocument} (HKDF) can generate a number of seeds $s_i$, to be used as inputs for deriving the indexes of the DIDs. That way, the node is required to store securely only $S$ and can regenerate the DIDs on the fly when needed during the operation phase, thus avoiding the secure storage of the entire set of DIDs. 

After the construction of the Merkle tree, each node interacts with a Trusted Party (TP). The identity key pair of the TP is $(sk_{TP}, pk_{TP})$. The TP provides the public key
$pk_{TP}$ to the node and the node shares the $ROOT$ of its Merkle tree with the TP.

Once the TP has collected all the $ROOT$ values from the $N$ nodes, it builds and publishes on the distributed ledger, at a given well-know predefined index $idx_{list}$, the list of trusted roots in the form: 
$$\{ROOT_1, \dots, ROOT_N, ROOT_{TP}; TS; Signature_{sk_{TP}} \}$$ 
where $TS$ is a timestamp that provides the date and time of list generation, and $Signature_{sk_{TP}}$ is the signature of the TP made with its private key $sk_{TP}$.

All nodes have simple access to this list of trusted roots by querying the DLT. The nodes
verify $Signature_{sk_{TP}}$ with the $pk_{TP}$ before consuming the list during the operation phase.

\subsection{Operation} 
\label{operation}
The nodes enter into digital interaction with the other nodes after (mutual) authentication. 
Upon the selection of a DID from the tree, a node generates the proof of membership that it will use during the authentication procedure with another node, as discussed in Section~\ref{intro}. 
The proof of membership coincides with the value of the Siblings of the corresponding leaf. For example, if the node $n_{1}$ selects $DID_1$, the proof is: $$(Sib_1, Sib_2) \doteq (H(idx_2),H(H(idx_3)|H(idx_4))).$$

When an interaction between node $n_1$ and a node $n_2$ takes place, first $n_2$ sends a nonce to $n_1$, meant to avoid replay attacks. Then $n_1$ sends $DID_{1}$, the proof of membership and a signature with its $sk_{id}$ on the message $H((Sib_1, Sib_2)|nonce)$ (\ie $Sig_{sk_{id}}(H((Sib_1, Sib_2)|nonce))$.

At that point, $n_2$ verifies the signature with the $pk_{id}$ retrieved from the DID Document pointed by $DID_1$, then it recalculates the root of the Merkle tree of node $n_{1}$ as:
$$ROOT_1 = H(H(H(idx_1)|Sib_1)|Sib_2).$$ 

The authentication succeeds if $ROOT_1$ is in the list of trusted roots, otherwise $n_{2}$ closes the communication. In case of mutual authentication, the same procedure takes place in both directions.

\subsection{Secret rotation}
Secret rotation is the process of replacing cryptographic secrets with new ones periodically or in response to a critical event. Key rotation is an example of this practice.

In case the TP needs to update its keys $(sk_{TP}, pk_{TP})$, the TP is required to provide to all nodes in the network its new public key $pk_{TP}'$ and, then, to sign and publish again the list of trusted roots on the distributed ledger at $idx_{list}$. 

Differently, when a node needs to update its identity keys ($sk_{id}, pk_{id}$), it selects a new DID in its Merkle tree, revokes the previous one, and generates and stores the new DID Document on the distributed ledger at the index pointed by the new DID. The proof of membership it will use during the authentication with another node changes accordingly, but since the $ROOT$ value does not change, the node does not need to interact with the TP. In other words, it updates the DID in full compliance with the SSI paradigm. Note that, identity key rotation does not necessarily imply the selection of a new DID, because W3C DID recommendation~\cite{DID} allows a node to update its DID Document without changing the DID.

A special case occurs when a node needs to update its identity keys ($sk_{id},pk_{id}$) and it is using the last DID in the Merkle tree (\eg\ $idx_4$ in Fig.~\ref{merkle}). 
The node autonomously generates a new Merkle tree and, then, shares the new $ROOT'$ value with the TP upon (mutual) authentication.  At that point, the node updates the DID, selecting the DID from the new tree, and the TP updates and publishes the new list of trusted roots. This key rotation process can continue indefinitely in normal operation phase in full compliance with the SSI paradigm. 

\subsection{Network update}

A new node can be deployed on the network without disrupting the operation of the other nodes. When the provisioning of the new node is concluded, the TP updates the list of the trusted roots and publishes it on the distributed ledger to inform the other nodes. 
The same happen when a node is removed from the network for any reason; the TP updates the list of trusted roots (\ie removes the corresponding $ROOT$ value) and publishes it on the distributed ledger.

\subsection{Critical analysis}

The following aspects are worth to be remarked: 

\begin{itemize}
    
    \item the solution is compliant with the SSI principles, since it does not affect the autonomy of a node to control its identity data ($sk_{id}$, DID, DID document);

    \item  the decentralized nature is respected; after the provisioning phase, a node follows the SSI paradigm without requiring major interactions with the TP; 
    
    \item  however, the role of the TP makes it a point of attack; the TP private key $sk_{TP}$ must be properly protected, because an adversary capable to gain access to $sk_{TP}$ can manipulate the list of trusted roots;    

    \item the solution scales with the number of nodes and it seems to be appropriate also for networks with high update frequency. Adding or removing a node only affects the size of the list of trusted roots  (\ie proportional to the number of active nodes) but it does not imply interactions with the other nodes in the network;

    \item a trade-off exists between the size of the Merkle tree, the size of the proof of membership (\ie the number of Siblings) and the number of interactions with the TP. The larger the tree size, the larger the proof, but the lower the number of interactions with the TP to send the root of a new Merkle tree;
       
    \item the security of the membership solution relies on the preimage resistance property of the hash function (\ie it is hard to invert) used to build the tree and on the HKDF used as a source of randomness to build the seeds; in this sense, it is reasonable to consider it quantum safe~\cite{mattsson2021};%
    
    \item both the Merkle tree and HKDF are mature technologies. However, their combined use in the proposed solution may need further security validation.        
    
\end{itemize}

\section{Membership through BBS Group Signature} 
\label{sect:BBS}
The BBS group signature scheme~\cite{BBS} has been developed to allow a member of the group to secretly sign a message on the group’s behalf with its group private key $gsk_{i}$; the signature of any member can be verified with the unique group public key $gpk$.  Moreover, the scheme allows a TP to revoke the private key of any member, triggering the update of the private keys of all the other members and of the group public key.

The BBS scheme is here adapted to solve the membership problem presented in Section~\ref{intro}. In principle, a node proves that its DID belongs to a trusted set by means of a BBS signature. The paper here adopts the notation from~\cite{BBS} and, when appropriate, directly refers to specific BBS algorithms, namely $KeyGen$, $Sign$, $Verify$, $Update$, $Open$, $Join$, and $Revoke$. 

\subsection{Provisioning} 
\label{BBSprovis}
The provisioning phase consists of the common configuration procedure in a secure environment of each node before deployment on the field. 

A TP supervises this phase and begins the provisioning by performing the $KeyGen$ algorithm. This step consists in the generation of the TP key pair ($sk_{TP}, pk_{TP}$), the group public key $gpk$, the TP group private key $tpsk$, and the private keys $gsk_{i}$ for all nodes in the network. 

The TP provides any node with its group private key $gsk_i$, the group public key $gpk$, 
the $index_{RL}$ on the distributed ledger where to retrieve a Revocation List,
and the public key $pk_{TP}$ to verify the TP signature on such list,
as will be explained in detail in Section~\ref{operationbbs}. 

According to the original $Join$ algorithm in \cite{BBS}, the TP generates and gives a group private key $gsk_i$ to each node; this protocol implies that the TP knows all the group private keys, making it a single point of attack. An alternative $Join$ algorithm, proposed in~\cite{ACJT}, introduces the property of \textit{Strong Exculpability} (SE) and, for clarity, will be denoted as $Join_{SE}$ in the following discussion.
The SE concept is an evolution of the exculpability concept that was first introduced by~\cite{AT}. In accordance with its definition in~\cite{BMW, ACJT}, SE ensures that no member of the group and not even the entity that issues the private keys can forge a signature on behalf of another group member. 

The authors of BBS in~\cite{BBS} suggested acquiring the SE property by generating each $gsk_{i}$ via a procedure in which the TP only learns a share of $gsk_{i}$. However, to the best of our knowledge, beside this suggestion, no practical implementation of the $Join_{SE}$ algorithm for BBS has been published. Therefore, we here propose for the first time an implementation tailored to the BBS group signature.

Firstly, the $KeyGen$ algorithm must be modified to add one more base to the original $gpk=(g_{1}, g_{2}, h, u, v, w)$ where $g_{1},\,h,\,u,\,v \in \mathbb{G}_1$ and $g_{2},\,w \in \mathbb{G}_2$ with $\mathbb{G}_1,\,\mathbb{G}_2$ multiplicative cyclic groups of prime order. Accordingly, the TP selects at random ${h_{1}}\overset{R}{\longleftarrow} \mathbb{G}_1$ and adds it to the new ${gpk} = (g_{1}, g_{2}, h, {h_{1}}, u, v, w)$.

Then, the $Join_{SE}$ algorithm can be constructed as follows:

\begin{enumerate}%

    \item the node $n_i$ selects at random $y_i  \overset{R}{\longleftarrow} \mathbb{Z}^{\ast}_{p}$ and sends $Y=h_{1}^{-y_{i}}$ to the TP;

    \item given $\gamma$ a TP's random secret value defined as $\gamma \in \mathbb{Z}^{\ast}_{p}$, the TP selects $x_{i} \overset{R}{\longleftarrow} \mathbb{Z}^{\ast}_{p}$,
    computes $A_{i} = (g_{1}Y)^{\frac{1}{\gamma+x_{i}}}$
    and $H_i = h_{1}^{\frac{1}{\gamma+x_{i}}}$, and sends $H_i$ to $n_i$; 
   
    \item $n_i$ sends $B_{i}=H_i^{-y_{i}}$ back to the TP;
    
    \item the TP computes $A_{i}'= B_{i}(g_{1}^{\frac{1}{\gamma+x_{i}}})$ and check if $A_{i}' = A_{i}$ to convince itself that $n_i$ knows $y_{i}$;
    
    \item if and only if %
          previous step 4) succeeds, the TP sends $(A_{i}, x_{i})$ to $n_i$;

    \item finally, $n_i$ builds its %
    entire private key $gsk_{i} = (A_{i}, x_{i}, y_{i})$.
\end{enumerate}

It is worth noting that only $n_i$ knows $y_{i}$. The Discrete Logarithm’s Problem (DLP) protects $y_{i}$ from being discovered by the TP that, in fact, only knows $Y=h_{1}^{-y_{i}}$ and $(A_{i},x_{i})$. In our solution $n_i$ proves the knowledge of the private key share $y_{i}$ in Zero-Knowledge. 

The $Sign$, $Verify$, $Update$, $Open$, $Join$, and $Revoke$ algorithms in~\cite{BBS} must be accordingly adapted to the new definition of $gsk_{i} = (A_{i}, x_{i}, y_{i})$. The adaptation consists in adding the Zero-Knowledge proof of knowledge of the entire group private key $gsk_{i}$, following the same approach used in constructing the $Join_{SE}$. These adaptations are here omitted for conciseness.

\subsection{Operation}
\label{operationbbs}
A node generates the DID and enters into interaction with other nodes of the network after proper (mutual) authentication. 
A node $n_1$ computes the proof of membership (\ie the BBS signature) running the $Sign$ algorithm using its $gsk_{1}$ on %
a digest computed as $H(DID_{n_{1}}|nonce)$, where the $nonce$ is generated by the counterpart node $n_2$ to avoid replay attacks. The %
node $n_2$ can verify the proof of membership with the $Verify$ algorithm using the group public key $gpk$ and, then, the ownership of $sk_{id}$ of $n_1$. In case of mutual authentication, the same procedure takes place in the two directions.

In any case, each node must maintain its own group private key and the group public key up to date. After the revocation of a key $gsk_r$, all nodes must update their own private key $gsk_i$ and $gpk$ with the $Update$ algorithm (see Sect. 7 in \cite{BBS}). 
The $Update$ algorithm requires some knowledge of the revoked private keys.
For this reason, the TP publishes a list of such knowledge, \ie a Revocation List (RL), 
on the distributed ledger at a well-know $index_{RL}$. All the nodes in the network can easily access this list by querying the distributed ledger.

According to the new $Join_{SE}$ algorithm, the %
RL contains a processed version of the share of the revoked private keys %
$(gsk^{\ast}_{r}, \dots, gsk^{\ast}_{s})$
known by the TP. %
The %
RL has the form: 
$$\{gsk^{\ast}_{r}, \dots, gsk^{\ast}_{s}; TS; Signature_{sk_{TP}} \}$$ 
where $TS$ is a timestamp that provides the date and time of the list, and $Signature_{sk_{TP}}$ is the signature of the TP. The nodes whose group private key $gsk_r$ is in the RL, cannot update their own $gsk_r$ by design of the $Update$ algorithm \cite{BBS}, hence they are no more able to generate a valid proof of membership.

\subsection{Secret rotation} 
 A node is able to update its identity keys $(sk_{id}, pk_{id})$ for key rotation purpose, and the respective DID and DID Document, in full compliance with SSI paradigm, without the need to update its $gsk_i$. The node must only generate the new proof when starting the authentication procedure with another node of the network. 

Differently, if the TP needs to update its keys $(sk_{TP}, pk_{TP})$, the TP has to share the new $pk_{TP}'$ with all the nodes, then sign with $sk_{TP}'$ and publish again the RL.

Moreover, if the TP needs to update its group secret key $tpsk$ for rotation purpose, the TP has to start a new provisioning phase to provide all the nodes with new group keys (\ie $gsk_{i}'$, $gpk'$). 

\subsection{Network update}
\label{Mandyngroup}
A new node can be deployed on the network without disrupting the operation of the other nodes. The TP concludes the $Join_{SE}$ procedure with the new node and, then, shares the group public key $gpk$, $index_{RL}$, and $pk_{TP}$ with it.

On the contrary, when a node is removed from the network for any reason, the TP performs the $Revoke$ algorithm and publishes the new RL. This revocation action causes all the other nodes to update their group keys. 

It is worth noting that the BBS group signature scheme provides a specific algorithm, named $Open$, that can be used to trace a signature to a signer (\ie retrieve a share of $gsk_{r}$ of the signer from a signature). This tools can be useful to detect a misbehaving node (\eg a compromised node) and revoke its group private key $gsk_{r}$.

\subsection{Critical Analysis}
The following aspects are worth to be remarked:
\begin{itemize}

    \item the solution is compliant with the SSI principles, since it does not affect the autonomy of a node to control its identity data ($sk_{id}$, DID, DID document);

    \item beside the provisioning phase where the TP provides the group keys to every node, the solution respects the decentralized nature of SSI;
   
    \item the revocation of a group private key implies some operations to be executed by all the other nodes (\ie they check the latest RL to update their group private keys $gsk_{i}$ and public key $gpk$ with the $Update$ algorithm);
        
    \item the TP can be identified as a single point of attack. Both private keys $sk_{TP}$ and $tpsk$ must be properly protected. An adversary gaining access to those secrets can add a malicious node to the network and revoke the capability of a legitimate node to make valid proofs of membership. The $Join_{SE}$ protocol offers a protection against the adversaries willing to generate a valid proof of membership on behalf of another node, since the TP does not know the full $gsk_{i}$; 

    \item the solution scales as the number of nodes increases. However, each revocation triggers the update of the group keys in all the other nodes. Notably, the size of the RL could grow with the number of revocations;

    \item the solution ensures total flexibility for the node to deal with its DIDs. In fact, once a node is provisioned with a valid $gsk_{i}$, it can freely create and update its DIDs, proving that they are in a trusted set by means of the BBS signature. Notably, the  signature has a constant size and there is not a trade-off between the dimension of the proof and other parameters of the solution;
    
    \item the security of the BBS group signature scheme relies on the Linear assumption and on the Strong Diffie-Hellman assumption. As a consequence, it can be considered vulnerable to attack by quantum computers \cite{Shor}; 
    
    \item finally, this solution is based on an already well-established and mature construction (\ie the BBS scheme), that can be used with minor modifications.%

\end{itemize}

\section{Performance estimation} 
\label{sect:perfo}

The feasibility of the two solutions is here addressed by estimating and comparing their computational load and expected performances on a target IoT node (\ie Raspberry Pi{\textsuperscript{\textregistered}} 4 Model B, 4 GB RAM, 1.5 GHz processor~\cite{rpi4brief}). 

This work adopts the same methodology applied in \cite{Canard2012} to estimate and to compare the execution time of the cryptographic operations in the four different operational phases.

First of all we have measured on the selected IoT node the execution time of the specific cryptographic algorithms heavily used as elemental building blocks in the two solutions under evaluation (\ie hash computation, scalar multiplication, exponentiation, and pairing). Table~\ref{tab:bench} shows the results of measurements assuming a 128-bit security level.

The initial benchmark shows that the \verb+sha256+ hash computation lasts 4 \textmu{s} and it is the less expensive cryptographic algorithm, whereas the paring computation lasts 50.4 ms and it is the most expensive one, as expected.
As an additional remark, the results in Table~\ref{tab:bench} are consistent with the values reported in \cite{Canard2012}, taking into account the different processor clock speed of the target nodes (\ie 1.5 GHz versus 1.2 GHz). 

These results are the basis for estimating and comparing the execution time of the two proposed solutions, as reported in the following subsections.

\subsection{Results for Merkle tree-based solution}

\begin{table}[ht]%
\caption{Benchmark results for %
specific cryptographic algorithms}
\begin{center}
\begin{tabular}{|l|c|c|}
    \hline 
    \textbf{Cryptographic Algorithm}  & \textbf{Notation} & \textbf{Time} (ms)\\
    \hline 
    Hash computation (\verb+sha256+)        & $\mathbf{h}$ & 0.004 \\
    Scalar multiplication in $\mathbb{G}_1$ & $\mathbf{m}$ & 4.6\\
    Exponentiation in $\mathbb{G}_T$        & $\mathbf{e}$ & 33.6\\
    Ate pairing $e$                         & $\mathbf{P}$ & 50.4\\ 
    \hline    
\end{tabular}
\label{tab:bench}
\end{center}
\vspace{5pt}

\caption{Estimated performance for Merkle tree-based solution}
\begin{center}
\begin{tabular}{|l|c|c|}
    \hline 
    \textbf{Operational Phase}  & \textbf{Estimated Computations} & \textbf{Time} (ms)\\
    \hline 
    \textit{Provisioning}       & $\mathbf{h}(4{k}+1)$ & 0.516\\ %
    \textit{Operation (Proof)}  & $\mathbf{h}(4{k}+1)$ & 0.516\\ %
    \textit{Operation (Verify)} & $\mathbf{h}(\log_2({k}) + 1)$  & 0.024 \\
    \textit{Secret Rotation}    & none or $\mathbf{h}(4{k}+1)$ & $\leq$ 0.516\\ %
    \textit{Network Update}     & none & 0 \\ 
    \hline
\end{tabular}
\label{tab:MTresults}
\end{center}
\vspace{5pt}

\caption{Estimated performance for BBS-based solution}
\begin{center}
\begin{tabular}{|l|c|c|}
    \hline 
    \textbf{Operational Phase}  & \textbf{Estimated Computations} & \textbf{Time} (ms)\\
    \hline 
    \textit{Provisioning}    & $2\mathbf{m}$ &  9.2\\ %
    \textit{Operation (Proof)}  & $\mathbf{h} + 5\mathbf{m} + 2(2\star\mathbf{m}) + 3\star\mathbf{e}$ & 75.7\\ %
    \textit{Operation (Verify)} & $\mathbf{h} + 4(2\star\mathbf{m}) + 4\star\mathbf{e}  + \mathbf{P}$ & 115.3\\ %
    \textit{Secret Rotation}   & none or $2\mathbf{m}$ & $\leq$9.2\\ %
    \textit{Network Update}    & none or $2\star\mathbf{m}$ & $\leq$5.4\\ %
    \hline
\end{tabular}
\label{tab:BBSresults}
\end{center}
\end{table}

The Merkle tree-based solution implies a computational load proportional to the size of the tree that, according to the structure in Fig.~\ref{merkle}, depends on the number of leaves on the tree (\ie DIDs).

Let us denote the number of leaves with $k$. This value is the key parameter to estimate the execution time in the four operational phases. In fact, it corresponds to the number of seeds $s_i$ to be generated by the HKDF and to the number of inputs to the hash algorithm to derive the indexes of the DIDs (\ie leaves of the Merkle tree). In addition, the number of leaves $k$ has an impact on the computational load to generate the Merkle tree, to create a proof of membership using the proper siblings and to verify a proof of membership given the siblings.

Table~\ref{tab:MTresults} reports the number of required computations and the estimated execution times for the selected IoT node, assuming a Merkle tree with 32 leaves (\ie ${k}=32$). These results neglect the operations executed by the TP and other operations with a limited impact on the computational load of the node (\eg random number generations). The rows of Table~\ref{tab:MTresults} represents the operational phases; it must be noted that the \textit{Operation} phase is split between the generation of a proof of membership (\ie \textit{Proof}) and its verification (\ie \textit{Verify}), since they can be executed by two distinct nodes (\ie $n_{1}$ and $n_{2}$, as explained in Section~\ref{operation}).

The generation of $k$ seeds $s_i$ with HKDF requires, according to \cite{HKDFdocument}, the following computations:

\begin{itemize}
    \item $2\mathbf{h}$ for the initial \verb+HKDF-Extract+ function, and 
    \item $2\mathbf{h}{k}$ to generate $k$ seeds with \verb+HKDF-Expand+ function.
\end{itemize}

Moreover, the generation of a Merkle tree from $k$ seeds requires $2{k}-1$ hash computations and, thus, a time equal to $\mathbf{h}(2{k}-1)$.

From these remarks, it is possible to state that the \textit{Provisioning} phase consists in the generation of $k$ seeds with HKDF plus the construction of a Merkle tree and, thus, it requires 
$2\mathbf{h} + 2\mathbf{h}{k} + \mathbf{h}(2{k}-1) = \mathbf{h}(4{k}+1)$. 

On the other hand, the verification of a proof of membership implies $\mathbf{h}(\log_2({k}) + 1)$ to compare the siblings against the $ROOT$ value.

Assuming that each node stores only a single master secret $S$ and regenerates the seeds $s_i$ and the Merkle tree on the fly when needed, thus avoiding the secure storage of the entire set of DIDs, the number of computations, hence the execution times, can be derived in the same way.

\subsection{Results for BBS-based solution}

The computation load for the BBS-based solutions has been estimated by considering the analytical results in \cite{Canard2012}. Table~\ref{tab:BBSresults} shows the number of required computations and
the estimated execution times for the selected IoT node.

This work considers all the optimizations suggested in \cite{Canard2012}, especially the general low level optimizations proposed in \cite{cheng2005implementing}, the optimal Ate pairing implementation in \cite{beuchat2010high}, and the suggestions in Section 6 of \cite{BBS} to eliminate all the pairings in the computation of a BSS signature and to perform only one pairing to verify a signature.

Moreover, the BBS scheme requires several multi-scalar multiplication and exponentiation operations, with a remarkable computational load.
For this reason, Table~\ref{tab:BBSresults} denotes a 
multi-scalar multiplication in $\mathbb{G}_1$ with 
$\ell \star \mathbf{m}$,
while a multi-scalar exponentiation in $\mathbb{G}_T$ is denoted as  
$\ell \star \mathbf{e}$,
where $\ell$ is the total number of multiplication or exponentiation operations to be executed.
For example, the second row of Table~\ref{tab:BBSresults} reports a  double scalar multiplication as $2\star\mathbf{m}$, while $3\star\mathbf{e}$ denotes a triple exponentiation.

This work considers also an optimization of these multi-scalar multiplication and exponentiation operations using a generalization of the Shamir's trick \cite{multiscalar} that, according to \cite{Canard2012}, allows accelerating these computations by a factor equal to 
$\frac{2^{\ell + 1} - 1}{3 \times 2^{\ell - 1}}$.

It must be noted that the results about the Operation phase (both Proof and verify) include a \verb+sha256+ digest computation that is executed before running the BBS $Sign$ and $Verify$ algorithms, respectively, according to Section~\ref{operationbbs}.

\section{Comparative analysis} 
\label{sect:comp}
The two proposed approaches show some similarities.
Both solutions take advantage of mature building blocks (\ie Merkle tree, HKDF, and BBS algorithms) 
and both comply with SSI principles (\ie they do not interfere with the decision of a node to create, update or revoke a DID, just add the mechanism to prove that the DID belongs to an evolving trusted set).
Moreover, both solutions respect the decentralized nature of SSI,
because, apart from the initial provisioning phase, they do not strictly require other direct interactions between the TP and the nodes. 

The TP is a single point of attack in both solutions, but with a difference. In the Merkle tree-based solution, an adversary capable to gain access to $sk_{TP}$ can arbitrarily compromise the list of trusted roots; in the BBS-based solution the adversary must gain access also to $tpsk$ to be able to add a malicious node to the network, or to revoke the capability of a legitimate node to make valid proofs of membership. In any case, a compromised TP has not direct access to the critical secrets of the nodes, especially their identity private keys $sk_{id}$.

The main difference between the two solutions resides in the provisioning phase. In the Merkle tree-based solution a node builds on its own the knowledge to generate the proof of membership (\ie the Merkle tree). In the BBS-based solution the TP generates and provides that knowledge to the node (\ie $gsk_i$). The $tpsk$ is the secret underpinning the group scheme. When the TP needs to update $tpsk$, for rotation purpose, the TP must start a new provisioning phase with all single nodes. In the former solution the nodes cyclically refresh their secrets (\ie the Merkle trees) and share the $ROOT$ values with the TP during the operation phase without interrupting their operation. Moreover, the adoption of a group signature scheme implies a less efficient revocation procedure, because it requires all nodes to update their group keys  $gsk_{i}$ and $gpk$ every time the TP revokes a group private key. However, apart from these %
disadvantages, the BBS-based solution provides full flexibility in DID creation, ensures a constant size for the proof that a DID belongs to a trusted set and does not impose any design constraint on the DID update. In fact, a node can potentially generate on the fly an unlimited number of DIDs, without the need to find a trade-off between the Merkle tree size and the number of interactions with the TP.

As far as the performance of the two solutions is concerned, the Merkle tree-based solution clearly outperforms the BBS-based solution in all the considered operational phases, especially in the Operation (Verify) and Network Update phases. It mainly relies only on fast hash computations and it does not require pairing computation at every verification or specific update operations at every revocation. 

For these reasons, the Merkle tree-based approach can be considered the most appropriate solution for IoT networks. On the other hand, the BBS-based solution could be of interest for possible use cases that require a constant/small size for the proof of membership in order to minimise the data exchange between nodes.

\section{Conclusions and future works} 
\label{sect:conclusion}
This paper has proposed an alternative (mutual) authentication process for a network of IoT nodes leveraging SSI. The main idea is to complement the DID-based verification of the identity private key ownership with the verification of a proof of membership during the (mutual) authentication process. 
The paper has analyzed two membership solutions, a novel solution designed by the Authors based on Merkle trees and a second solution built as an adaptation of the BBS group signature scheme.
The performance evaluation has provided an estimate of the computational load on an IoT node for each method, while the comparative analysis has highlighted the advantages and drawbacks of both solutions.

Future works will focus on (\emph{i}) the adoption of threshold signature schemes to reduce the impact of a possible attack to the TP, and (\emph{ii}) the adoption of dynamic accumulators and their properties to build another possible alternative.

\section*{Acknowledgment}

The Authors would like to thank Alberto Carelli for his technical support to the performance estimation of the two proposed solutions. 

\bibliographystyle{./IEEEtran}
\bibliography{Biblio}

\end{document}